\documentclass[epj,twocolumn]{webofc}
\usepackage[varg]{txfonts}   
\woctitle{Hadron Collider Physics symposium 2012}
\begin{document}
\title{Single top quark measurements with CMS}
%
%

\author{Jeannine Wagner-Kuhr\inst{1}\fnsep\thanks{\email{Jeannine.Wagner-Kuhr@kit.edu}}\\
on behalf of the CMS Collaboration 
}

\institute{Karlsruhe Institute of Technology (KIT), Germany
          }

\abstract{%
Measurements of single top quark production are presented, performed using CMS data collected in 2011 and 2012 at centre-of-mass energies of 7 and 8 TeV. The cross sections for the electroweak production of single top quarks in the $t$-channel and in association with W-bosons is measured and the results are used to place constraints on the CKM matrix element $V_{tb}$. 
}
\maketitle
\section{Introduction}
\label{intro}
Besides the dominant production of top quarks in pairs through the strong interaction top quarks can also be produced singly through charged-current electroweak interactions. Due to the large top-quark mass, these processes are well suited to test the predictions of the standard model (SM) of particle physics and to search for new phenomena. Measurements of the single top quark production cross section also provide a direct determination of the magnitude of the Cabibbo-Kobayashi-Maskawa (CKM) matrix element $|V_{tb}|$.

Single top quark production was observed in proton-antiproton collisions at the Tevatron collider with a centre-of-mass energy of 1.96 TeV~\cite{STdiscovery}. The cross section increases by a factor of 20 (27) at the Large Hadron Collider (LHC) running at $\sqrt{s}=7\,\mbox{TeV}$ ($\sqrt{s}=8\,\mbox{TeV}$) with respect to the Tevatron. 

In the standard model three types of electroweak single top quark production channels exist: t-channel and s-channel processes, and W-associated single top quark production (tW). The dominant contribution to the cross section is expected to be from the t-channel process. While at the Tevatron tW production is negligible, it gives the second largest contribution to the cross section at the LHC. The s-channel production cross section is very small at the LHC compared to the other two single top channels. At approximate NLO and for a top-quark mass of $m_t = 172.5\,\mbox{GeV/c}^2$, a $t$-channel production cross section of $\sigma_{t-ch.}^{th,7TeV}  = 66.2^{+2.6}_{-2.0}\;\mbox{pb}$ at $\sqrt{s}=7\,\mbox{TeV}$ and of $\sigma_{t-ch.}^{th,8TeV}  = 87.6^{+3.3}_{-2.4}\;\mbox{pb}$ at $\sqrt{s}=8\,\mbox{TeV}$, respectively, is predicted, while for W-associated single top quark production a cross section of $\sigma_{tW}^{th,7TeV}  = 15.7^{+1.2}_{-1.2}\;\mbox{pb}$ at $\sqrt{s}=7\,\mbox{TeV}$ and of $\sigma_{tW}^{th,8TeV}  = 22.4^{+1.6}_{-1.5}\;\mbox{pb}$ at $\sqrt{s}=8\,\mbox{TeV}$, respectively, is predicted~\cite{Kidonakis}.

As the single top quark cross section is proportional to the square of the CKM matrix element $|V_{tb}|$ the measurement of the single top quark cross section allows a direct measurement of $|V_{tb}|$. Furthermore, single top quark production is sensitive to new physics, like anomalous couplings, additional heavy quarks, heavy bosons like W' or the charged Higgs. In the following recent results from CMS~\cite{CMS} on the $t$-channel and tW production are presented.
\section{t-channel production}
\label{sec-tchan}
At $\sqrt{s}=7\,\mbox{TeV}$ the CMS collaboration has three $t$-channel cross section analyses~\cite{CMS-tchan7}, one relatively simple analysis, the $|\eta_{j'}|$-analysis, and two multivariate analyses, the BDT and NN analysis. The $|\eta_{j'}|$-analysis exploits the reconstructed top mass and the $t$-channel specific angular distribution of the light quark recoil jet $j'$, which goes in the forward/backward direction. The most critical backgrounds are modeled in a data-driven way. Hence, this analysis is straightforward, robust and has little model dependence. In the second approach, maximum precision using multivariate techniques is aimed for. First, the signal to background discrimination is optimized by exploiting the full event characteristics predicted by the SM. Secondly, systematic uncertainties are constrained by simultaneously analysing regions with substantial signal content and regions with negligible signal. As this approach is highly complex, two independent analyses have been conducted. After checking that the results of the three analyses are consistent with each other all three results are combined using the best linear unbiased estimator, BLUE~\cite{blue}, to get a slightly improved final result.

The $|\eta_{j'}|$-analysis has also been used to measure the $t$-channel cross section at $\sqrt{s}=8\,\mbox{TeV}$~\cite{CMS-tchan8}. Due to the higher center of mass energy and due to the more severe pile-up conditions some adjustments had to be made when moving from 7 TeV to 8 TeV.

\subsection{Event selection}
In all analyses $t$-channel events, where the W-boson decays leptonically in either a muon or an electron and the corresponding neutrino are considered. The events are triggered with a single muon trigger and an electron+b-jet trigger, respectively. Exactly one isolated muon or electron in the central region and with a transverse momentum $p_t$ above 20 and 30 GeV/c is required, respectively. Due to this isolation requirement multijet QCD background is strongly reduced. Then, at least two reconstructed jets with $p_t$ above 30 GeV, whereof at least one jet is tagged as b-jet are required. This b-tagging requirement reduces the background from W+light events strongly and after this cut W+$b\bar{b}$ background is one of the most important backgrounds. To further reduce multijet QCD background a cut on either the transverse mass of the W-boson or the missing transverse enery is applied. In the $|\eta_{j'}|$-analysis two additional cuts are applied. Here, exactly two jets are required whereof one jet is b-tagged and the reconstructed top quark mass has to be in a window around the true top quark mass: $130\,\mbox{GeV/c}^2<m_{\ell\nu b}<220\,\mbox{GeV/c}^2$. In the 8 TeV $|\eta_{j'}|$-analysis some cuts have been hardened due to the more severe pile-up conditions. Furthermore, as top quark pair events are the largest background in the 8 TeV analysis, not only W+jets and multijet QCD events but also top quark pair events are modeled in a data-driven way.

\subsection{Signal and background modeling}
The signal, single top quark $t$-channel events, as well as also the other single top quark processes are modeled using the NLO Monte Carlo (MC) generator POWHEG~\cite{powheg} and Pythia~\cite{pythia} for the hadronization. W+jets, Z+jets and top quark pair events are modeled using Madgraph~\cite{madgraph} MC together with Pythia and dibosn events are modeled with Pythia. Multijet QCD background is in all analyses modeled with data. The shape is obtained from data with a loose lepton selection, and the normalization is obtained from a fit to either the transverse mass distribution of the W-boson or to the missing transverse energy. The $|\eta_{j'}|$-analysis extracts the W+jets event yield and the shape from the data in the sideband of the reconstructed top quark mass. In this sideband region, single top quark events are negeligible. To get the W+jets yield and shape other backgrounds like top quark pair and multijet QCD production are subtracted. In the 8 TeV analysis, top quark pair production is the dominant background. To deal with this the $|\eta_{j'}|$-analysis uses a sideband region strongly enriched with top pair events, namely 3-jet events where two jets are tagged as b-jets, to model the shape of top pair production.

\subsection{Signal extraction}
\begin{table*}
\centering
\caption{Measured $t$-channel production cross section at 7 TeV and 8 TeV.}
\label{tab-tchan1}       
\begin{tabular}{ll}
\hline
 & $t$-channel cross section [pb]  \\\hline
\textbf{7 TeV} & \\
$|\eta_{j'}|$ & $70.0\pm 6.0\,\mbox{(stat.)}\pm 6.5\,\mbox{(syst.)}\pm 3.6\,\mbox{(theor.)}\pm 1.5\,\mbox{(lumi.)}$\\[0.1cm]
NN & $68.1\pm 4.1\,\mbox{(stat.)}\pm 3.4\,\mbox{(syst.)}^{+ 3.3}_{-4.3}\,\mbox{(theor.)}\pm 1.5\,\mbox{(lumi.)}$\\[0.1cm]
BDT & $66.6\pm 4.0\,\mbox{(stat.)}\pm 3.3\,\mbox{(syst.)}^{+ 3.9}_{-3.3}\,\mbox{(theor.)}\pm 1.5\,\mbox{(lumi.)}$\\[0.1cm]\hline
Comb. & $67.2\pm 3.7\,\mbox{(stat.)}\pm 3.0\,\mbox{(syst.)}\pm 3.5\,\mbox{(theor.)}\pm 1.5\,\mbox{(lumi.)}$\\[0.1cm]\hline\hline
\textbf{8 TeV} & \\
$|\eta_{j'}|$ & $80.1\pm 5.7\,\mbox{(stat.)}\pm 11.0\,\mbox{(syst.)}\pm 4.0\,\mbox{(lumi.)}$\\[0.1cm] \hline
\end{tabular}
\end{table*}
In the multivariate (MV) analyses categories are defined according to the number of reconstructed jets and the number of identified b-jets. Categories with substantial signal content (2-jets, 1-tag and 3-jets, 1-tag) as well as with negligible singnal content are defined. The W+light enriched categories (0-tag categories) are used to check the modeling of the W+jets background and it has been checked that not only the input variables to the MV analyses yield good agreement between data and MC but also the MV discriminators itself. In the fully Bayesian statistical inference the $t$-channel signal strength is extracted simultaneously by using either the BDT or NN discriminator output in the two signal enriched categories and the categories enriched in top quark pair production (2-jets, 2-tags; 3-jets, 2-tags; 4-jets, 1/2-tags) and simultaneously for the muon and electron channel. This enhances the signal acceptance and allows to constrain systematic uncertainties in situ.

In the $|\eta_{j'}|$-analysis the $t$-channel signal strength is obtained from a likelihood fit to the sensitive variable $|\eta_{j'}|$. The largest uncertainties in all three 7 TeV analyses are due to the W+jets modeling, due to b-tagging and due to theoretical uncertainties. Summing up the indidual uncertainties a total relative uncertainty of about 10\% is expected for the multivariate analyses.

\begin{figure}
\centering
\includegraphics[width=7.0cm]{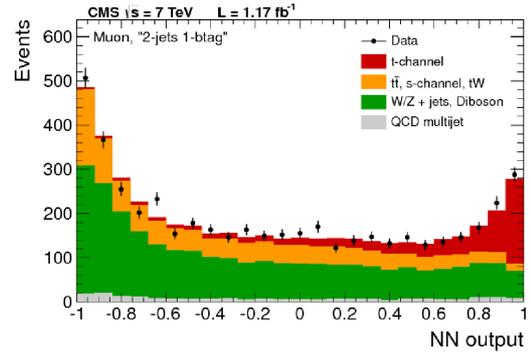}
\caption{NN discriminator output distribution of the 7 TeV analysis in the 2-jets, 1-tag category after scaling the singal and the background contributions to the best fit result.}
\label{fig-tchan1}       
\end{figure}
\begin{figure}
\centering
\includegraphics[width=7.0cm]{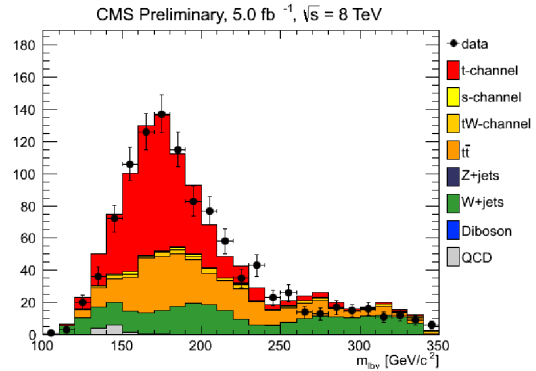}
\caption{Reconstructed top quark mass of the 8 TeV analysis after applying a cut on the recoil jet ($|\eta_{j'}|>2.0$) to enhance the signal purity.}
\label{fig-tchan2}       
\end{figure}
In Figure~\ref{fig-tchan1} as an example the NN discriminator is presented in the 2-jets, 1-tag category of the 7 TeV analysis after scaling the singal and the background contributions to the best fit result. It is clearly visible, that without $t$-channel single top quark production the prediction would not describe the data. The reconstructed top quark mass of the 8 TeV analysis after applying a cut on the recoil jet ($|\eta_{j'}|>2.0$) to enhance the signal purity is shown in Figure~\ref{fig-tchan2}. As one would expect for single top quark events, a prominent peak around the top quark mass is visible.

\subsection{Results}
\begin{figure}
\centering
\includegraphics[width=7.0cm]{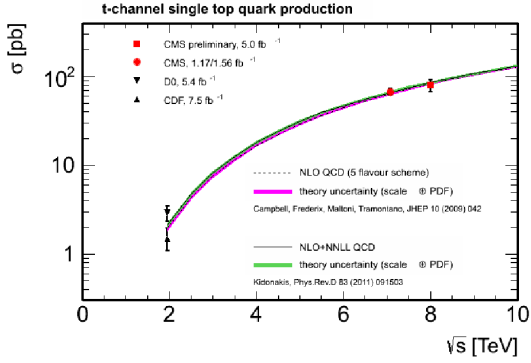}
\caption{$t$-channel cross section as a function of the center of mass energy.}
\label{fig-tchan3}       
\end{figure}

The measured $t$-channel production cross sections at 7 TeV and 8 TeV are given in Table~\ref{tab-tchan1}. The measured cross sections at 7 TeV of the three analyses are well consistent with each other and the combination using BLUE yields a cross section of $\sigma_{t-ch.}^{7TeV}=67.2\pm 6.1\,\mbox{pb}$ with a relative uncertainty of 9.1\%. This measured cross section is in good agreement with the SM prediction of $\sigma_{t-ch.}^{th,7TeV}  = 66.2^{+2.6}_{-2.0}\;\mbox{pb}$. At 8 TeV a cross section of $\sigma_{t-ch.}^{8TeV}=80.1\pm 13.0\,\mbox{pb}$ is measured which is also well consistent with the SM prediction. All the measured $t$-channel single top quark cross sections from CMS are summarized in Figure~\ref{fig-tchan3}, where the cross section as a function of the center of mass energy is presented.

The measured single top cross section is then used to measure the CKM matrix element $|V_{tb}|$. Assuming $|V_{td}|$, $|V_{ts}| \ll |V_{tb}|$ yields $|f_{L_V}V_{tb}|=\sqrt{\sigma_{t-ch.}/\sigma_{t-ch.}^{th}}=1.020\pm 0.046\,\mbox{(exp.)}\pm 0.017\,\mbox{(theor.)}$, where $f_{L_V}$ is the left-handed vector coupling at the Wtb vertex. Assuming for $f_{L_V}$ the SM value of 1 and exploiting a Feldman-Cousins approach yields the confidence interval $0.92<|V_{tb}|\le 1$ at 95\% C.L..

\section{W-associated single top quark production}
\label{sec-tW}

Using 7 TeV data, two analyses to search for W-associated single top quark production (tW) have been performed by the CMS collaboration~\cite{CMS-tW}. The reference analysis is a multivariate analysis using Boosted Decision trees and the second analysis is a simple and robust cut-based analysis to give confidence in the result of the more sophisticated MV analysis.

\subsection{Event selection}
Both tW analyses use the dilepton channel, where the two W-bosons decay leptonically in muons or electrons. To trigger the events dilepton triggers are used
and exactly two isolated leptons in the central region and with $p_t > 20\,\mbox{GeV/c}$ are required. Cuts on the invariant dilepton mass are applied to reject low invariant mass Drell-Yan events and events with a real Z-boson. To further reduce the background from Z+jets events (but also from multijet QCD events) a cut on the missing transverse energy is applied in the $ee$ and $\mu\mu$ channels. Furthermore, one or two reconstructed central jets with $p_t > 30\,\mbox{GeV/c}$ are required, whereof at least one jet is b-tagged. To reduce top quark pair background, events that have additional low $p_t$ b-tagged jets are rejected. In the simple cut-based analysis one additional cut is applied. Here, the scalar sum of the transverse momenta of the two leptons, the jets, and the missing transverse energy is required to be above 60 GeV in the $e\mu$-channel, where no missing transverse energy cut is applied. After all these cuts the dominant background is top quark pair production

\subsection{Modeling of signal and backgrounds}
Single top quark and background processes are generated as described for the $t$-channel analyses.

Due to the interence of tW and top quark pair production the definition of tW poses conceptual issues and in this analysis the diagram removal (DR) scheme to define tW is used. In the DR scheme all NLO tW diagrams that are doubly resonant are excluded from the tW definition. Differences of the DR scheme and the alternative diagram subtraction (DS) scheme, in which the differential cross section is modified with a gauge-invariant subtraction term, that locally cancels the contribution of top quark pair production, are accounted for in the systematic uncertainties.

In studies it was found that in high-pileup scenarios the missing transverse energy for Z events is not properly modeled by the simulation. To deal with this missing transverse energy dependent data-to-MC scale factors are determined in events from the Z resonance and are then applied to the selected Z+jets MC events. In both analyses a signal region and two control regions strongly enhanced with top quark pair events are used. The signal region contains events with exactly one b-tagged jet, while the top quark pair control regions contain events with two jets whereof either one or both are b-tagged. By perfomring a simultaneous fit to the data in the signal and in the top quark pair control regions top quark pair background is constrained.

\subsection{Signal extraction} 
In general there is no single variable that separates well between top-quark pair production and tW production. To increase the separation power in the signal region four variables that have some ability to separate between top quark pair and tW production are combined in a BDT. Those variables are $H_T$, defined as the scalar sum of the leptons, jet, and missing transverse energy, the transverse momentum of the reconstructed tW-system, the $p_t$ of the leading jet, and the azimuthal angle between the missing transverse energy vector and the closest lepton.
\begin{figure}[htb]
\centering
\includegraphics[width=7.0cm]{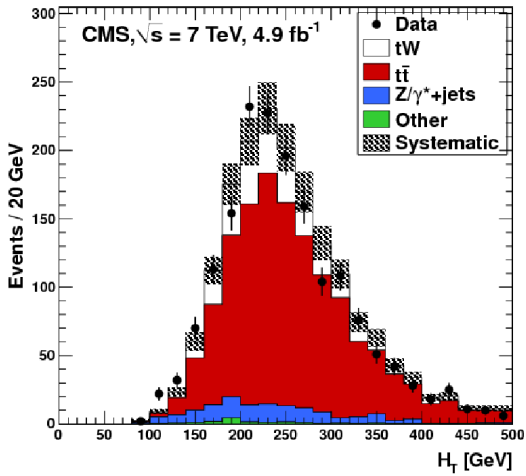}
\caption{ $H_T$, the scalar sum of the leptons, jet and missing transverse energy, in the signal region (1-jet, 1-tag).}
\label{fig-tW1}       
\end{figure}

\begin{figure}[htb]
\centering
\includegraphics[width=7.0cm]{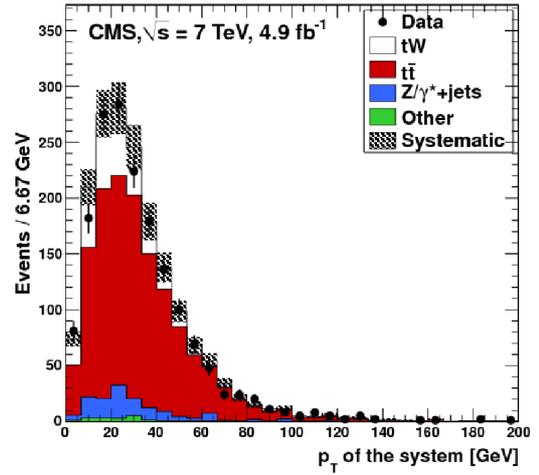}
\caption{Transverse momentum of the reconstructed tW-system in the signal region (1-jet, 1-tag).}
\label{fig-tW2}       
\end{figure}

Figure~\ref{fig-tW1} shows $H_T$ and Figure~\ref{fig-tW2} the transverse momentum of the tW-system in the singal region. Here, the MC prediction is scaled to the luminosity of the data. Good agreement between data and MC is obtained.

In the BDT analysis, a simultaneous fit to the BDT outputs of the three lepton channels in the signal region and to the yield in the top-quark pair control regions of the three lepton channels is performed. In case of the cut-based analysis only the rates of the signal and the two control regions are used in the statistical inference. In both analyses systematic uncertainties are constrained insitu. The largest uncertainties are here due to the uncertainty of the jet energy scale and due to theoretical uncertainties.

\subsection{Results}
\begin{table*}
\centering
\caption{Expected and observed significance for tW-roduction together with the measured tW cross section at 7 TeV.}
\label{tab-tW1}       
\begin{tabular}{lll}
\hline
 & BDT-analysis & cut-based analysis  \\\hline
Expected sensitivity & $3.6^{+0.8}_{-0.6}\,\sigma$ & $3.2\pm 0.9\, \sigma$ \\
Observed Sensitivity & $4.0\,\sigma$ & $3.5\,\sigma$ \\
Measured cross section & $16^{+5}_{-4}\,\mbox{pb}$ & $15\pm 5\,\mbox{pb}$\\ \hline
\end{tabular}
\end{table*}

\begin{figure}[h]
\centering
\includegraphics[width=7.0cm]{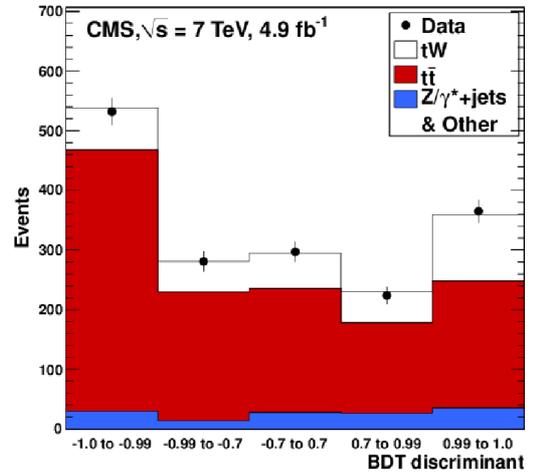}
\caption{BDT putput in the signal region (1-jet, 1-tag) after scaling the signal and background compositions to the best fit result.}
\label{fig-tW3}       
\end{figure}
In Figure~\ref{fig-tW3} the BDT output distribution in the signal region is presented after scaling the signal and background contributions to the best fit result. It is clearly visible, that without tW-production the data would not be well described by the prediction. Table~\ref{tab-tW1} summarizes the results of the two analyses using 7 TeV data. The expected sensitivity of both analyses is above $3\,\sigma$. The observed significance of the BDT analysis is $4.0\,\sigma$, while for the first time also a simple cut-based analysis yields an observed significance of above $3\,\sigma$. The measured cross sections for W-associated single top quark production are in good agreement with the approximate NNLO SM prediction of $\sigma_{tW}^{th,7TeV}  = 15.7^{+1.2}_{-1.2}\;\mbox{pb}$.

Also the measured tW cross section can be used to determine the CKM matrix element $|V_{tb}|$. The result of $|V_{tb}|=\sqrt{\sigma_{tW}/\sigma_{tW}^{th}}=1.01^{+0.16}_{-0.13}\,\mbox{(exp.)}^{+0.03}_{-0.04}\,\mbox{(theor.)}$ is with the current statistics not as precise as the one obtained in the $t$-channel but it is well consistent with the SM value very close to 1 and with the dermination of $V_{tb}$ in the t-channel.

\section{Summary}
\label{sec-summary}
In this article recent results from CMS on $t$-channel and W-associated single top quark production are presented. All measurements are within the current statistics in good agreement with the SM prediction. 

Concerning the t-channel, the cross section is measured with a relative precision slightly better than 10\%, and $|V_{tb}|$ has been determined under some assumptions with a relative precision of 5\%. This shows that the LHC experiments enter with single top quark production the era of precision and properties measurement including of course also the searches for new physics.

Concerning W-associated single top production evidence for this channel has been found with $4\,\sigma$ using a multivariate analysis and this evidence has been confirmed with a simple cut-based analysis. The next step will be the ultimate establishment of this channel and more detailed measurements.


\begin{thebibliography}{}
%
%
\bibitem{STdiscovery}
CDF Collaboration, Phys. Rev. \textbf{D82}, 112005 (2010); D0 Collaboration, Phys. Rev. \textbf{D84}, 112001 (2010).
\bibitem{Kidonakis}
Nikolaos Kidonakis, arXiv:1210.783 (2012); private communication.
\bibitem{CMS}
CMS Collaboration, JINST \textbf{03}, S08004 (2008).
\bibitem{CMS-tchan7}
CMS Collaboration, JHEP \textbf{1212}, 035 (2012).
\bibitem{blue}
J. Pumplin, D. R. Stump, J. Huston et al., JHEP \textbf{07}, 012 (2002).
\bibitem{CMS-tchan8}
CMS Collaboration, CMS PAS-TOP-11-021 (2012).
\bibitem{powheg}
E. Re, Eur. Phys. J. C \textbf{71}, 1547 (2011);
S. Alioli, P. Nason, C. Oleari et al., JHEP \textbf{06}, 043 (2010);
S. Alioli, P. Nason, C. Oleari et al., JHEP \textbf{09}, 111 (2009);
S. Frixione, P. Nason, and C. Oleari, JHEP \textbf{11}, 070 (2007).
\bibitem{pythia}
T. Sj\"ostrand, S. Mrenna, and P. Z. Skands, JHEP \textbf{05}, 026 (2006).
\bibitem{madgraph}
J. Alwall, M. Herquet, F. Maltoni et al., JHEP \textbf{06}, 128 (2011).
\bibitem{CMS-tW}
CMS Collaboration, Phys. Rev. Lett. \textbf{110}, 022003 (2013).
\end{thebibliography}
\end{document}